\documentclass[10pt]{article}

\usepackage{graphicx}

\usepackage{amsmath}
\usepackage{multirow}
\usepackage{amssymb}

\newcommand{\twotwo}
[4]{\left(\begin{array}{cc} #1 & #2 \\ #3 & #4\end{array}\right)}
\newcommand{\cZ}{{\mathcal Z}}
\newcommand{\cY}{{\mathcal Y}}
\newcommand{\cX}{{\mathcal X}}
\newcommand{\W}[1]{\langle W_{#1}|}
\newcommand{\V}[1]{|V_{#1}\,\rangle}
\newcommand{\cA}{{\mathcal A}}

\newcommand{\cT}{{\mathcal T}}
\newcommand{\bbZ}{{\mathbb Z}}
\newcommand{\bbN}{{\mathbb N}}

\begin{document}

\title{Exclusive Queueing Process with Discrete Time}

\author{Chikashi Arita\thanks{
Faculty of Mathematics, Kyushu University}
\ and Daichi Yanagisawa\thanks{
Department of Aeronautics and Astronautics, School of Engineering, The University of Tokyo, and
Japan Society for the Promotion of Science (JSPS)
}
}

\date{ }

\maketitle

\begin{abstract}
In a recent study \cite{RefA},
 an extension of the M/M/1 queueing process
 with the excluded-volume effect
 as in the totally asymmetric simple exclusion process
 (TASEP) was introduced.
In this paper, we consider its discrete-time version.
The update scheme we take is the parallel one.
A stationary-state solution is obtained
 in a slightly arranged matrix product form
 of the discrete-time open TASEP with the parallel update.
We find the phase diagram for the existence of the stationary state.
The critical line which separates the parameter space 
 into regions with and without the stationary state can be written in terms of
 the stationary current of the open TASEP.
We calculate the average length of the system
 and the average number of particles.
\end{abstract}

\section{Introduction} \label{intro}

The queueing theory has been considerably studied since Erlang started designing telephone exchanging system in 1909 \cite{ake1}. 
In \cite{ake2}, he developed the theory of the call-loss system,
 which significantly contributed to the progress of telephones and electric communication systems.
After his study, Kendall presented his first paper \cite{dgk1}.
Since then, the study on the queueing theory
 including Kendall's notation \cite{dgk2}, Burke's theorem \cite{pjb},
 Jackson networks \cite{jkj}, and Little's theorem \cite{little}
 has been accelerated.
Nowadays, it is applied to study
 on social systems such as the Internet \cite{q&c1},
 resource management systems \cite{q&c0},
 vehicular traffic systems \cite{q&p1}
 and pedestrian traffic systems \cite{q&p0}.

Let us review here the discrete-time M/M/1 queueing process,
 which will be extended in the next section.
\begin{figure}\begin{center}
\centering
\includegraphics{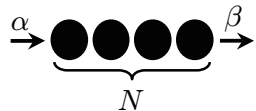}
\caption{M/M/1 queueing process.}
\label{fig:mm1q}
\end{center}\end{figure}
M/M/1 means that the queue has
   a Markovian entry of customers
   and a Markovian exit at 1 server.
Each particle (customer) enters the system with a probability $\alpha$
 and receives service with a probability $\beta$,
 see figure \ref{fig:mm1q}.
The system is characterized
 only by the number $N\in \bbZ_{\ge0}$ of particles.
The probability $P(N;t)$ of finding $N$ particles
 at time $t$ is governed by the following master equation:
\begin{align}
 P(0;t+1)
 =& (1-\alpha) P(0;t) + \alpha\beta P(0;t)
 +  (1-\alpha)\beta P(1;t), \\
\begin{split}
 P(N;t+1)
 =& \alpha(1-\beta) P(N-1;t)+(1-\alpha)(1-\beta) P(N;t) \\
  &+  \alpha\beta P(N;t) +  (1-\alpha)\beta P(N+1;t),
\end{split}
\end{align}
for $N\in\bbN$.
When $\alpha<\beta$, 
 a unique stationary-state solution
 to the master equation can be easily obtained
 as the following geometric distribution:
\begin{align}
P(N)= 
\frac{\beta-\alpha}{\beta(1-\alpha)}
\left(\frac{\alpha(1-\beta)}{(1-\alpha)\beta}\right)^N.
\end{align}
The average number of particles
 can be easily calculated as
\begin{align}
 \langle N\rangle =\sum_{N=1}^{\infty}
     N\frac{\beta-\alpha}{\beta(1-\alpha)}
     \left(\frac{\alpha(1-\beta)}{(1-\alpha)\beta}\right)^N
  =\frac{\alpha(1-\beta)}{\beta-\alpha}.
\end{align}

On the other hand,
 the asymmetric simple exclusion process (ASEP)
 in one dimension is one of typical interacting particle systems
 and admits exact analyses of non-equilibrium properties
 \cite{RefL,RefS}.
It has been vigorously studied in the recent two decades.
The stationary state
 of the totally ASEP with open boundaries (open TASEP)
 was found in a matrix product form in \cite{RefDEHP}.
Since then, matrix product stationary states of
 various generalizations of the ASEP
 containing discrete-time versions of the open TASEP
 have been found \cite{RefBE}.
The TASEP is one of the basic models
 of the vehicular traffic \cite{C&S&S}.

Although both the M/M/1 queueing process and the open TASEP
 are Markov processes with particle entry and exit,
 they are different in the following two points.
First, the number of particles in the M/M/1 queueing process
 does not have an upper limit\footnote
 {Queueing processes with an upper limit
 of the number of particles have been studied.\cite{RefJMB}},
 whereas that in the open TASEP cannot be greater than
 the fixed number of sites.
In other words,
 the state space of the M/M/1 queueing process is infinite
 and that of the open TASEP is finite.
It should be noticed that the open TASEP is a call-loss system,
 which has been seldom discussed.
Second,
 the TASEP has a spatial structure,
 and particles are affected
 by their excluded volume (hard-core repulsion).
By contrast, the M/M/1 queueing process has no spatial structure,
 and only the number of particles characterizes the system.

\begin{figure}\begin{center}
\centering
\includegraphics{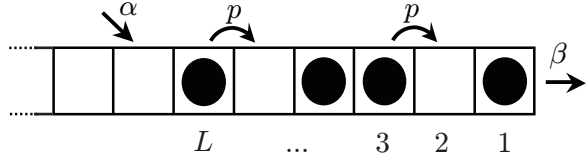}
\caption{Exclusive queueing process.}
\label{fig:eq}       
\end{center}\end{figure}

An extension of the M/M/1 queueing process
 with the excluded-volume effect
 (in other words, the TASEP with a new boundary condition)
 was introduced in \cite{RefA}.
In this model,
 particles enter the system
 at the left site next to the leftmost occupied site
 and exit at the rightmost site of the chain,
 see Fig. \ref{fig:eq}.
When we (pedestrians) make a queue,
 we usually proceed if there is a space in front of us
 (excluded-volume effect),
 which is a motivation to treat this model.
Let us call the model {\it exclusive queueing process}.

In \cite{RefA}, a stationary state of the exclusive queueing process
 was given by a slightly arranged matrix product form for the open TASEP,
 where the probability of finding each configuration
 is proportional to
\begin{align}\label{arrangedMPF}
 c^L \times \langle\text{row vector}|
 \text{ matrix product }
 |\text{column vector}\rangle .
\end{align}
The row vector, the matrices and the column vector
 are independent of the entry rate $\alpha$, whereas
 $c$ is independent of the exit rate $\beta$.
The exponent $L$ denotes the position of the leftmost particle.

In this paper, we consider a discrete-time version of
 the exclusive queueing process
 with the entry, hopping and exit rates
 ($\alpha, p$ and $\beta$, respectively)
 replaced by probabilities within one time step.
Although there are some choices of an update scheme,
 we take the fully-parallel-update rule.\footnote
{The discrete-time open TASEPs
 with the random-, forward-sequential-, backward-sequential-,
 sub-lattice-parallel- and fully-parallel-update
 schemes have been studied, see \cite{RefBE} for review.
It is reported that the movement of pedestrians in one dimension
 is well represented by the
 parallel-update dynamics \cite{CR-AS-AS-TGF09}.
} 
The special case where the bulk hopping is deterministic
 (i.e., the hopping probability is 1)
 has been analyzed in \cite{dy04jsiaml}.

To find a stationary-state solution is one of basic problems.
The idea to find it in this paper is reducing
 the balance equation for the exclusive queueing process
 to that for the open TASEP.
Then, a stationary-state solution will be obtained
 in a similar form to that for
 the continuous-time model as \eqref{arrangedMPF}.
We calculate the critical line which separates the parameter space
 into the regions with and without the stationary state
 by evaluating the convergence of the normalization constant.
The critical line will be written
 in terms of the stationary current of the open TASEP.
We calculate the average length of the system
 (the position of the leftmost particle) and
 the average number of particles in the system.
We also calculate the average number of particles in the queue.\footnote
{We define the number of particles in the queue
by counting particles except for one at the rightmost site.}

This paper is organized as follows.
In Sect. \ref{model}, we define
the exclusive queueing process with discrete time.
In Sect. \ref{stat-state},
 we obtain a stationary state of the model.
In Sect. \ref{pdav},
 we identify the region where the stationary state
 exists in the parameter space,
 and calculate the average length of the system
 and the average number of particles.
In Sect. \ref{sec:p=1}, we treat the model in the case where the bulk hopping probability is 1.
Section \ref{conc} is devoted to the conclusion of this paper.
Detailed calculations are done in Appendices.

\section{Model} \label{model}

\begin{figure}\begin{center}
\centering
\includegraphics[height=57mm,clip]{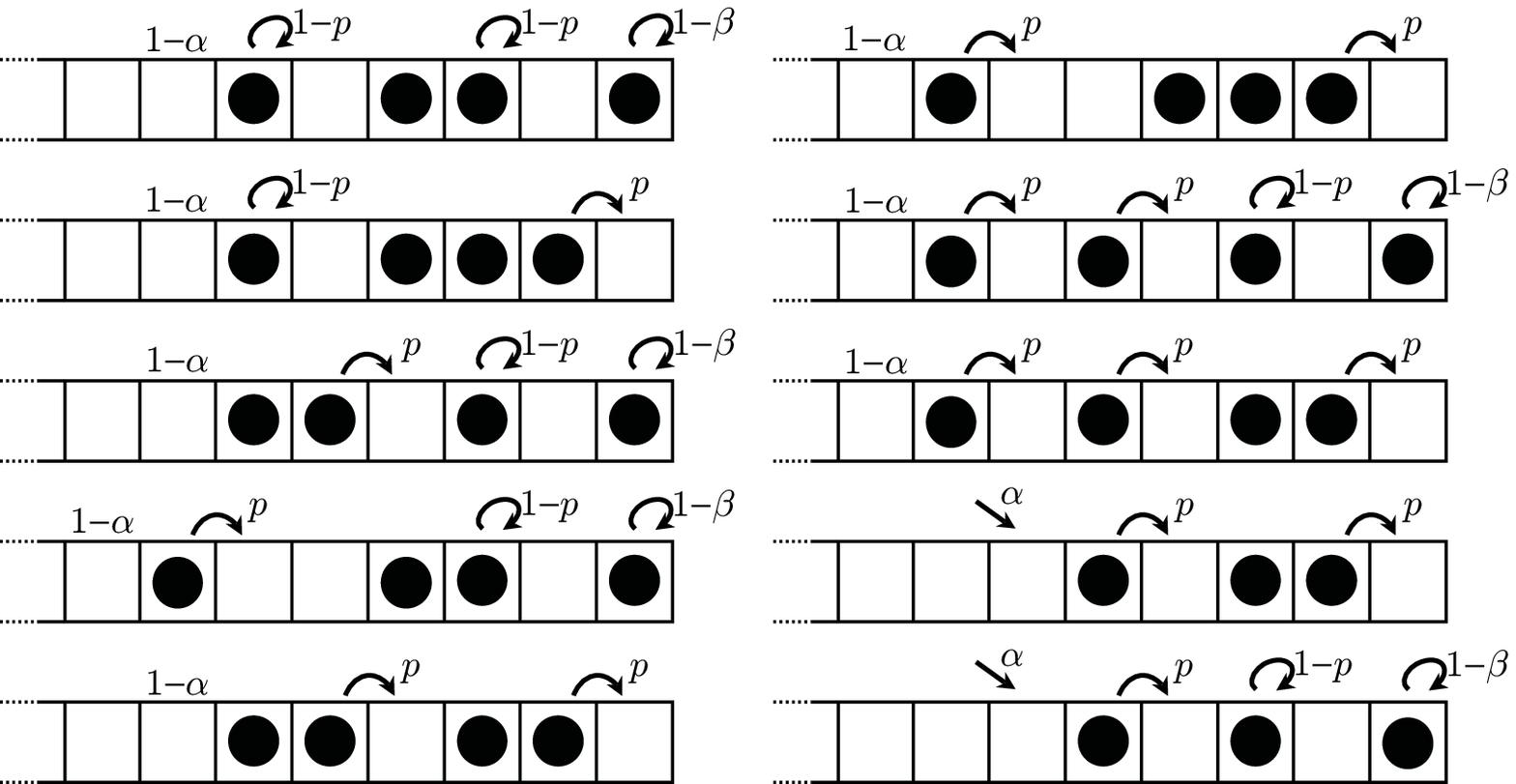}
\caption{List of all configurations which transit to $101101$
  with a non-zero probability within one time step.}
\label{fig:101101}
\end{center}\end{figure}

\begin{figure}\begin{center}
\centering
\includegraphics[height=45mm,clip]{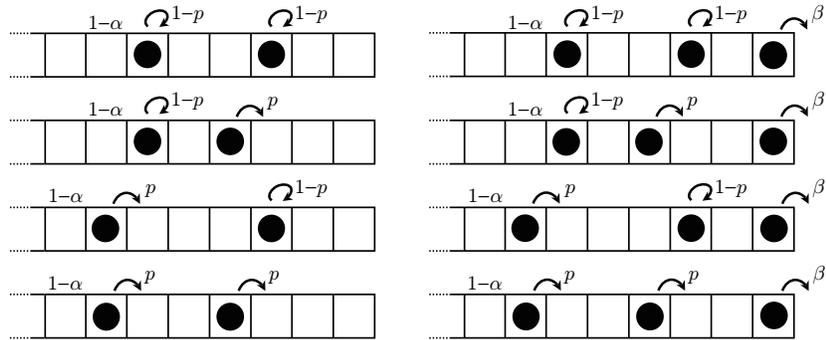}
\caption{List of all configurations which transit to $100100$
  with a non-zero probability within one time step.}
\label{fig:100100}
\end{center}\end{figure}

We introduce an extension of the discrete-time M/M/1 queueing process
 with the excluded-volume effect on the semi-infinite chain
 (exclusive queueing process),
 see Fig. \ref{fig:eq},
 where each site is labeled by a natural number from right to left.
Each particle enters the chain
 at the left site next to the leftmost occupied site
 with a probability $\alpha$,
 hops to its right nearest neighbor site
 with a probability $p$, if it is empty,
 and exits at the right end of the chain
 with a probability $\beta$.
If there is no particle on the chain, a particle enters at site 1.
These transitions occur simultaneously within one time step.
In other words, we take the fully-parallel-update scheme.
(We call it simply parallel update hereafter.)
The model is formulated as a discrete-time Markov process
 on the state space $S=\{\emptyset,1,10,11,100,101,110,111,1000,\dots\}$
 where 0 and 1 correspond to unoccupied and occupied sites, respectively.
In particular, $\emptyset$ denotes the state
 that there is no particle on the chain. 
Note that we do not write infinite 0s
 located left to the leftmost 1.
We define $|\tau|$ by the length of a sequence $\tau$.
In particular, for each element $\tau$ of $S$,
 $|\tau|$ stands for the length of the system
 (the position of the leftmost occupied site).
Let us write the probability of finding a configuration $\tau$
 at time $t$ as $P(\tau;t)$.
The master equation governing the model is written as
\begin{align}
 P(\tau;t+1)=\cT P(\tau;t)
 =\sum_{\tau'\in S} W(\tau'\to\tau)P(\tau';t),
\end{align}
where $\cT$ is the transition-probability matrix of the process
 and $W(\tau'\to\tau)$ is the transition probability
 from $\tau'$ to $\tau$ within one time step.
In particular, for $\tau=\emptyset$ and $\tau=1$, 
\begin{align}
 & P(\emptyset;t+1) =  (1-\alpha) P(\emptyset;t) + (1-\alpha)\beta P(1;t), \\
 & P(1;t+1) = \alpha P(\emptyset;t)+(1-\alpha)(1-\beta) P(1;t)
               + (1-\alpha)p P(10;t).
\end{align}
It is difficult to write down the action of $\cT$ explicitly for the general configuration $\tau$
 because of the parallel update\footnote{
It is not so difficult in the continuous-time case,
 see (22) in \cite{RefA}.}.
For example, for $\tau=101101$,
\begin{align}
\begin{split}
 &\!\!\!\!\!\! P(101101;t+1) \\
 =&(1-\alpha)(1-p)^2(1-\beta) P(101101;t) 
  +(1-\alpha)(1-p)p P(101110;t) \\
 &+(1-\alpha)p(1-p)(1-\beta) P(110101;t) \\
 &+(1-\alpha)p(1-p)(1-\beta)P(1001101;t) 
  +(1-\alpha)p^2 P(110110;t) \\
 &+(1-\alpha)p^2 P(1001110;t)
 +(1-\alpha)p^2(1-p)(1-\beta) P(1010101;t) \\
 &+(1-\alpha)p^3 P(1010110;t)
  +   \alpha p^2 P(10110;t) \\
 &+   \alpha p(1-p)(1-\beta)P(10101;t),
\end{split}
\end{align}
see Fig. \ref{fig:101101}, and for $\tau=100100$,
\begin{align}
\begin{split}
 &\!\!\!\!\!\! P(100100;t+1) \\
 =&(1-\alpha)(1-p)^2 P(100100;t)
  +(1-\alpha)(1-p)p P(101000;t) \\
 &+(1-\alpha)p(1-p) P(1000100;t) 
 +(1-\alpha)p^2 P(1001000;t) \\
 &+(1-\alpha)(1-p)^2\beta P(100101;t)
  +(1-\alpha)(1-p)p\beta P(101001;t) \\
 &+(1-\alpha)p(1-p)\beta P(1000101;t) 
 +(1-\alpha)p^2\beta P(1001001;t),
\end{split}
\end{align}
see Fig. \ref{fig:100100}.

\begin{figure}\begin{center}
\centering
\includegraphics{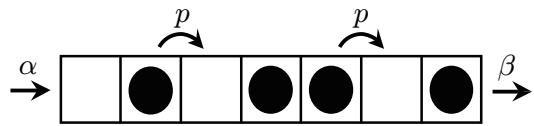}
\caption{TASEP with the ordinary open boundary condition.}
\label{fig:opentasep}       
\end{center}\end{figure}

Note that this model is not equivalent to the
 parallel-update TASEP with the ordinary open boundary condition (open TASEP) \cite{RefERS},
 see Fig. \ref{fig:opentasep}.
In the open TASEP, particles can enter the system
  at the fixed leftmost site of the finite chain only if it is empty.
This means that the open TASEP is a call-loss system.
On the other hand, in our model, particles can always enter the system.

\section{Stationary state} \label{stat-state}

A stationary state (stationary distribution) is a solution to
 the balance equation
\begin{align}\label{balance}
P(\tau)=\cT P(\tau)\ (\forall \tau\in S)
\end{align}
 with the constraint
\begin{align}\label{constraint}
 \sum_{\tau\in S}P(\tau)=1.
\end{align}
Let us begin with making an assumption
 that the stationary-state solution has the following form:
\begin{align}\label{ass1}
 P(\emptyset)=&\frac{1}{Z}, \\
 P(1\tau_{L-1}\dots\tau_1)
 =&\frac{1}{Z}\left( \frac{\alpha}{p(1-\alpha)} \right)^L
   f_L(1\tau_{L-1}\dots\tau_1),
\label{ass2}
\end{align}
where $f_L$ is a function on $\{0,1\}^L$ and satisfies
\begin{align}
\label{f1}
 f_L(101\tau)&=  f_{L-1}(11\tau) + pf_{L-2}(1\tau),\\
\label{f2}
 f_L(100\tau)&= (1-p)f_{L-1}(10\tau), \\
\label{f3}
 f_{L}(01\tau)&= pf_{L-1}(1\tau), \\
\label{f4}
 f_L(00\tau)&=0.
\end{align}

From $P(\emptyset)=\cT P(\emptyset)$, we get 
\begin{align}
 P(1)=\frac{1}{Z}\frac{\alpha}{p(1-\alpha)}\frac{p}{\beta}.
\end{align}
From $P(1)=\cT P(1)$, we get
\begin{align}
 P(10)=\frac{1}{Z} \left(\frac{\alpha}{p(1-\alpha)}\right)^2
 \frac{p}{\beta}.
\end{align}
Thus,
\begin{align}
 \label{f11} f_1(1)=\frac{p}{\beta},\\ 
 \label{f210} f_2(10)=\frac{p}{\beta}.
\end{align}
We reduce the balance equation for the general configuration
 to that for the open TASEP.
One can obtain, for all $11\tau\in S$ with length $L$,
\begin{align}
\begin{split}
\label{11}
f_L(11\tau)
=& \sum_{\tau'\in\cA_{11}(\tau)}
    \frac{1}{1-\alpha}W(11\tau'\to11\tau) f_L(11\tau') \\
 &+\sum_{\tau'\in\cA_{11}(\tau)}
    \frac{1}{\alpha}W(1\tau' \to11\tau) f_L(01\tau'),
\end{split}
\end{align}
 for all $101\tau\in S$ with length $L$,
\begin{align}
\begin{split}
\label{101}
f_L(101\tau)
=& \sum_{\tau'\in\cA_{101}(\tau)}
  \frac{1}{1-\alpha}W(101\tau'\to101\tau)f_{L}(101\tau') \\
 &+\sum_{\tau'\in\cA_{101}(\tau)}
  \frac{1}{1-\alpha}W(110\tau'\to101\tau)f_{L}(110\tau') \\
 &+\sum_{\tau'\in\cA_{101}(\tau)}
  \frac{1}{\alpha}W(10\tau'\to101\tau)f_{L}(010\tau'),
\end{split}
\end{align}
and for all $100\tau \in S$ with length $L$,
\begin{align}
\begin{split}
\label{100}
f_L(100\tau)=&  \sum_{\tau'\in\cA_{100}(\tau)}
 \frac{1}{1-\alpha}W(10\tau'\to100\tau)f_L(10\tau') \\
 & +\sum_{\tau'\in\cA_{100}(\tau)}
   \frac{1}{(1-\alpha)(1-p)}W(10\tau'\to100\tau)f_L(00\tau'),
\end{split}
\end{align}
see Appendix \ref{appa} for detailed calculation.
The sets $\cA_{\dots}(\tau)$ are defined as
\begin{align}
\cA_{11}(\tau):=&\{\tau'| W(11\tau'\to 11\tau)>0
   \wedge|\tau|=|\tau'|\}, \\
\cA_{101}(\tau):=&\{\tau'| W(101\tau'\to 101\tau)>0
   \wedge|\tau|=|\tau'|\}, \\
\cA_{100}(\tau):=&\{\tau'| W(10\tau'\to 100\tau)>0
   \wedge|\tau|=|\tau'|-1\}.
\end{align}
For example,
\begin{align}
 \cA_{11}(0110) &= \{ 0110,1010,0111,1011 \}, \\
 \cA_{101}(110) &= \{ 110,111 \}, \\
 \cA_{100}(110) &= \{ 0110,1010,0111,1011 \}.
\end{align}
Equations \eqref{11}-\eqref{100}
 give the balance equation for the open TASEP on the $L$-site chain
 with the entry probability equal to 1
 if we regard $f_L(\tau)$ as the probability of finding a configuration $\tau$.
Thus, if we set $f_L$ to be proportional to the stationary-state solution
 to the open TASEP,
 the form \eqref{ass1} and \eqref{ass2} gives
 a stationary-state solution to our model.
Substituting the matrix product solution to the open TASEP
 which was found in \cite{RefERS} into \eqref{ass2},
 we obtain
\begin{align}\label{arrmpf}
 P(\tau)=
\begin{cases}
   \frac{1}{Z} & \tau=\emptyset,   \\
   \frac{1}{Z}\left(\frac{\alpha}{p(1-\alpha)} \right)^L
   \langle W | DX_{\tau_{L-1}}\cdots X_{\tau_1} |V\rangle        & \tau=1\tau_{L-1}\dots\tau_1,
\end{cases}
\end{align}
where $X_1=D$ and $X_0=E$ are matrices, 
$\W{}$ is a row vector and $\V{}$ is a column vector.
They should satisfy 
\begin{align}\label{algrel}
\begin{split}
  EDEE =& (1-p)EDE + EEE + pEE, \\
  EDED =& EDD + EED + pED,\\
  DDEE =& (1-p)DDE + (1-p)DEE + p(1-p)DE, \\
  DDED =& DDD + (1-p)DED + pDD, \\
 DDE\V{} =& (1-\beta)DD\V{} + (1-p)DE\V{} + p(1-\beta)D\V{} ,  \\
 EDE\V{} =& (1-\beta)ED\V{} + EE\V{} + pE\V{}, \\
 \W{} DEE =&  (1-p) \W{} DE, \\
 \W{} DED =& \W{} DD + p\W{} D, \\
 DD\V{} =& \frac{p(1-\beta)}{\beta}D\V{},  \\
 ED\V{} =& \frac{p}{\beta}E\V{}, \\
 \W{} EE =&  0, \\
 \W{} ED =& p\W{} D,
\end{split}
\end{align}
which is the algebraic relation found in \cite{RefERS}
  with $\alpha=1$.
This algebraic relation
  is compatible with the assumption \eqref{f1}-\eqref{f4}.
A representation of the relation is given in Appendix \ref{appb}.
In view of \eqref{f11}, we have to chose the normalization as
\begin{align}\label{WDV}
 \W{} D \V{} = \frac{p}{\beta}.
\end{align}
Note that, from the algebraic relation, we can derive $\W{} DE \V{} = \frac{p}{\beta}$,
 which is compatible with \eqref{f210}.
The normalization constant $Z$
 is expressed as
\begin{align}\label{norcon}
 Z =1+\sum_{L\ge 1} \left(\frac{\alpha}{p(1-\alpha)}\right)^L
  \W{} D(D+E)^{L-1}\V{}
\end{align}
when the right-hand side converges.
This form is similar to the generating function
 of the normalization constant of the open TASEP.
The function
\begin{align}\label{series}
 \cZ (\xi,\zeta)
 =& 1 + \sum_{L\ge 1} \xi^L\langle W| \zeta D (\zeta D+E)^{L-1} | V \rangle
\end{align}
is useful to calculate the average length of the system
 and the average number of particles.
The case where $\xi=\frac{\alpha}{p(1-\alpha)}$ and $\zeta=1$
 corresponds to the normalization constant:
\begin{align}
 Z = \cZ \left(\frac{\alpha}{p(1-\alpha)},1\right).
\end{align}

We can easily solve the balance equation
  of the usual M/M/1 queue recursively.
(The equation is just a three-term recurrence formula.)
In our model, however,
  we have not found a recursive way to solve the balance equation.
In this sense,
   our assumption \eqref{ass1} and \eqref{ass2} is truly an Ansatz
   (or working hypothesis).

In the continuous-time case,
  a similar form gives its stationary state
  except that
  $\frac{\alpha}{1-\alpha}$ is replaced by $\alpha$,
  and an overall constant $\frac{p}{\beta}$ appears
  instead that
  the first $D$ in the matrix product disappears,
  see (25) in \cite{RefA}.
However, the continuous-time limit
  (first replace $\alpha\to\alpha\Delta t,
  \beta\to\beta\Delta t$ and $p\to p\Delta t$,
  and then take the limit $\Delta t\to 0$)
  of the solution \eqref{arrmpf}
  is exactly the same as the solution
  to the continuous-time model.

At this stage, we do not know if
  the power series for $Z$
   (the right-hand side of \eqref{norcon}) converges
   and the solution \eqref{arrmpf} is meaningful.
Note that the form \eqref{arrmpf}
   without the normalization constant always gives
   a stationary measure of our model,
   which is a solution to the balance equation \eqref{balance}
   without the constraint \eqref{constraint}.
If a process on a countable state space
  is irreducible and recurrent,
  and has two stationary measures $\mu$ and $\nu$,     
  then $\mu$ is proportional to $\nu$ (Proposition II.1.3 of \cite{RefSi}).
Furthermore, if an irreducible process has a stationary state,
  the process is (positive) recurrent
  and thus its stationary state is unique
 (Theorem II.1.1 of \cite{RefSi}).
Our process is easily seen to be irreducible
  for generic $\alpha,\beta$ and $p$.
Thus the stationary state \eqref{arrmpf}
   is unique if the power series for $Z$ converges
   and there is no (normalizable) stationary state if it diverges
   (Corollary II.1.2 of \cite{RefSi}).
In the next section,
  we obtain the condition on the parameters such that
  the power series for $Z$ converges
  and a more explicit closed form for $Z$.

\section{Phase diagram and average values}\label{pdav}

We can derive the following closed form for $\cZ(\xi,\zeta)$,
 see Appendix \ref{appb} for details:
\begin{align}\label{closedform}
 \cZ(\xi,\zeta)=\frac{1}{1-\frac{p}{\beta} \cY(\xi,\zeta)},
\end{align}
where
\begin{align}
 \cY(\xi,\zeta)=&
 \frac{\cX(\xi,\zeta)-\xi(1-\zeta )
   -\sqrt{(\cX(\xi,\zeta)-\xi(1-\zeta ))^2-4 \xi\zeta\cX(\xi,\zeta)}}
   {2(1+p\xi\zeta)}, \\
 \cX(\xi,\zeta)=& (1+p\xi)(1+p\xi\zeta).
\end{align}
The power series for $\cZ(\xi,1)$
 (the right-hand side of \eqref{series} with $\zeta=1$)
 converges to
$\frac{1}{1-\frac{p}{\beta}\frac{(1+p\xi)-\sqrt{(1+p\xi)^2-4\xi}}{2}}$
  when 
\begin{align}
 \begin{cases}
  \xi \le \left(\frac{1-\sqrt{1-p}}{p}\right)^2 & \beta>1-\sqrt{1-p},\\
  \xi  < \frac{(p-\beta)\beta}{p^2(1-\beta)} & \beta\le1-\sqrt{1-p}.
 \end{cases}
\end{align}
Replacing $\xi$ by $\frac{\alpha}{p(1-\alpha)}$,
 we obtain the region
 where the power series for $Z$
 (the right-hand side of \eqref{norcon}) converges as
\begin{align}\label{region}
\begin{cases} 
 \alpha\le\alpha_c=\frac{1-\sqrt{1-p}}{2} & \beta>1-\sqrt{1-p}, \\
 \alpha<\alpha_c=\frac{\beta(p-\beta)}{p-\beta^2} & \beta\le 1-\sqrt{1-p},
\end{cases}
\end{align}
and its closed form as
\begin{align}\label{Zclosed}
  Z = \frac{2(1-\alpha)\beta}{R-p+2(1-\alpha)\beta},
\end{align}
where $R=\sqrt{p(p-4\alpha(1-\alpha))}$.

\begin{figure}\begin{center}
\centering
\includegraphics[height=60mm,clip]{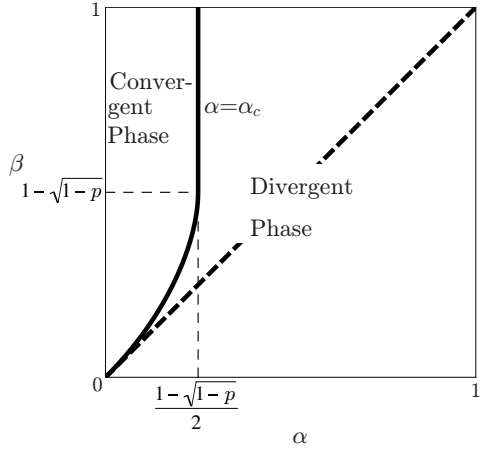}
\caption{Phase diagram for the parallel-update
exclusive queueing process.
The critical line $\alpha=\alpha_c$ (solid line)
  divides the parameter space into
  the convergent phase and the divergent phase.
We drew the dashed line representing the critical line
  of the usual M/M/1 queueing process as well.
Note that there is a region $\alpha_c<({\rm or}\le)\alpha<\beta$
  where the M/M/1 queueing process converges
  but our new model diverges.
}
\label{fig:pdgen}
\end{center}\end{figure}

As we see Fig. \ref{fig:pdgen},
 the region \eqref{region} is embedded in
 the region  $\alpha<\beta$ where the usual M/M/1 queueing process converges.
The critical line $\alpha=\alpha_c$ consists
 of two parts; a curve and a straight line.
Mathematically,
 the curve is given by a solution to
 the denominator of the form \eqref{Zclosed} $=0$,
 and the straight line by a solution to $R=0$.
Physically,
 the two parts correspond to the stationary current of
 the open TASEP in the high-density phase
 ($\beta<\min (1-\sqrt{1-p},\alpha)$)
 and the maximal-current phase
 ($\alpha,\beta\ge 1-\sqrt{1-p}$), respectively,
 see Table \ref{tab:disccont}.
This property is due to the form \eqref{norcon}.
In fact, the stationary current of the
 open TASEP on $L$-site chain with the entry probability 1
 is given by
\begin{align}
 J_L=\frac{pc_L}{c_{L+1}+pc_L},
\end{align}
where $c_L$ is the coefficient of $\xi^L$ in $\cZ(\xi,1)$.
Note that the critical line for the continuous-time
 version of the model can be also written
 in terms of the stationary current of the continuous-time open TASEP 
\cite{RefA},
see Table \ref{tab:disccont} again.

\begin{table}
\caption{
Comparison with the continuous-time model.
In the second row, the settings of the models are given.
In the third row, 
 the stationary currents of the open TASEPs
 in the limit where the length of the chain $L\to\infty$
 are described.
In the fourth row, 
 the critical lines of the exclusive queueing processes
 are described.
The right column is obtained
 by taking the continuous-time limit of the left column.
}
\label{tab:disccont}
\begin{tabular}{ll}
\hline\noalign{\smallskip}
discrete time (parallel update) & continuous time  \\
\noalign{\smallskip}\hline\noalign{\smallskip}
entry probability $\alpha$ & entry rate $\alpha$ \\
exit probability $\beta$      & exit rate $\beta$ \\
hopping probability $p$       & hopping rate $p$ \\
\noalign{\smallskip}\hline\noalign{\smallskip}
$J_{\infty}$ & $J_{\infty}$ \\
$
    =\begin{cases}
       \displaystyle \frac{1-\sqrt{1-p}}{2} & \alpha,\beta\ge 1-\sqrt{1-p} \\
       \displaystyle \frac{\alpha(p-\alpha)}{p-\alpha^2} & \alpha\le\min (1-\sqrt{1-p},\beta) \\
       \displaystyle \frac{\beta(p-\beta)}{p-\beta^2} & \beta<\min (1-\sqrt{1-p},\alpha) 
     \end{cases} $
 & $
   =\begin{cases}
       \displaystyle \frac{p}{4}      & \alpha,\beta\ge\frac{p}{2} \\
       \displaystyle \alpha(1-\alpha/p) & \alpha\le\min (\frac{p}{2},\beta) \\
     \displaystyle \beta(1-\beta/p) & \beta<\min (\frac{p}{2},\alpha) 
     \end{cases}$ \\ 
\noalign{\smallskip}\hline\noalign{\smallskip}
$\alpha$ & $\alpha$ \\
$
    =\begin{cases}
       \displaystyle \frac{1-\sqrt{1-p}}{2}
    & \beta\ge 1-\sqrt{1-p} \\
     \displaystyle \frac{\beta(p-\beta)}{p-\beta^2} &\beta<1-\sqrt{1-p}
     \end{cases} $
 & $
   =\begin{cases}
       \displaystyle \frac{p}{4}      & \beta\ge\frac{p}{2} \\
     \displaystyle \beta(1-\beta/p) & \beta<\frac{p}{2}
     \end{cases}$ \\ 
\noalign{\smallskip}\hline\noalign{\smallskip}
\end{tabular}
\end{table}

Let us calculate some average values in the stationary state.
We can calculate the average length of the system
 (the position of the leftmost particle)
 and the average number of particles in the system
 by differentiating $\ln\cZ(\xi,\zeta)$
 with respect to $\xi$ and $\zeta$, respectively, as
\begin{align}
 \langle L\rangle=&
 \xi\frac{\partial}{\partial \xi}\ln\cZ (\xi,1)
 \bigg|_{\xi=\frac{\alpha}{p(1-\alpha)}}
 =\frac{\alpha p(R-p+2(1-\alpha))}{R(R-p+2(1-\alpha)\beta)}, \\
 \langle N\rangle=&
 \frac{\partial}{\partial \zeta}
 \ln\cZ \left(\frac{\alpha}{p(1-\alpha)},\zeta \right)
 \bigg|_{\zeta=1}
 =\frac{\alpha(1-\alpha) (p-2\alpha p+R)}{R(R-p+2(1-\alpha)\beta)}.
\end{align}

The number $N_q$ of particles in the queue
 (i.e., $N_q=\#\{j\ge2|\tau_j=1\}$
  for $\tau_L\tau_{L-1}\cdots\tau_1\in S$ and
   $N_q=0$ for $\emptyset$)	
 is also one of the typical values in the queueing theory.
We can calculate the average of $N_q$ as
\begin{align}
 \langle N_q\rangle
 =&\frac{1}{Z}\frac{\partial}{\partial\zeta}
  \sum_{L\ge 2} \xi^L
  \W{} \zeta D (\zeta D+E)^{L-2} (D+E) \V{}
  \bigg|_{\xi=\frac{\alpha}{p(1-\alpha)},\zeta=1} \\
 =&\frac{1}{Z}\frac{\partial}{\partial\zeta}
    \left[\left(1+\frac{p\xi(1-\zeta)}{\beta(1+p\xi\zeta)}\right)
    \cZ(\xi,\zeta)-\frac{\beta+p\xi}{\beta}
  \right]_{\xi=\frac{\alpha}{p(1-\alpha)},\zeta=1} \\
 =&\langle N\rangle-\frac{\alpha}{\beta}.
\end{align}
Derivation of the second equality is given in Appendix \ref{appc}.

\section{Case $p=1$}\label{sec:p=1}

In this section we treat the model
 with $p=1$, where 
 the bulk hopping rule is nothing but
 one of the typical deterministic cellular automata,
 i.e., rule 184.
This case was analyzed in \cite{dy04jsiaml}.
The matrices and vectors have a two-dimensional representation 
\begin{align}
 D = \twotwo{\frac{1-\beta}{\beta}}{0}{ \frac{1}{\sqrt{\beta}} }{0},\ 
 E = \twotwo{0}{\frac{1}{\sqrt{\beta}} }{0}{0},\ 
 \langle W |= \left(1,\sqrt{\beta} \right),\ 
 |V\rangle = \left(\begin{array}{c}1\\ \sqrt{\beta} \end{array}\right),
\end{align}
see \cite{RefRSSS}.

In the case where $p=1$,
 each particle necessarily hops if its right neighbor site is empty.
Thus, any configurations containing at least one sequence 00
 vanish in the stationary state.\footnote{
This means that the process is no longer irreducible when $p=1$.
But the process on the subset of the state space $S$
   consisting of $\emptyset$ and all the configurations
   without the sequence 00 is irreducible.
}
In fact, we can see
\begin{align}
P(1\tau_{L-1}\dots00\dots\tau_1)
=\frac{1}{Z}\left(\frac{\alpha}{1-\alpha}\right)^L\W{}\cdots EE\cdots\V{}=0.
\end{align}
One can reduce the stationary-state probability
 to the following simpler expression:
 for $\tau=\tau_L\tau_{L-1}\cdots\tau_1$ ($\tau_L=1$),
\begin{align}
 P(\tau)
 =\begin{cases}
   0 & {\rm if}\ \tau \ {\rm contains\ a\  sequence\  00} \\
   \frac{1}{Z}\left(\frac{\alpha}{1-\alpha}\right)^{L}
   \frac{(1-\beta)^{2\# \{j | \tau_j=1\}-L-\tau_1 }}
{\beta^{\#\{j | \tau_j=1\} }}  
  &  {\rm otherwise}
 \end{cases}
\end{align}

\begin{figure}	
$\!\!\!\!\!\!\!\!\!\!\!\!\!\!\!\!\!\!\!\!\!\!\!\!\!\! 
 \!\!\!\!\!\!\!\!\!\!\!\!\!\!\!\!\!\!\!\!\!\!\!\!\!\!$
\includegraphics[height=65mm,clip]{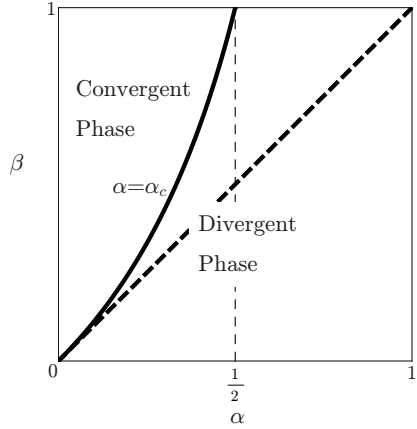}
\caption{Phase diagram for $p=1$.
The critical line $\alpha=\alpha_c$ (solid line)
  divides the parameter space into
  the convergent phase and the divergent phase.
We drew the dashed line representing the critical line
  of the usual M/M/1 queueing process as well.
Note that there is a region $\alpha_c\le\alpha<\beta$
  where the M/M/1 queueing process converges
  but our new model diverges.}
\label{fig:pd184}
\end{figure}

Thanks to the two dimensional representation,
 calculating $\cZ(\xi,\zeta)$ is an easy exercise:
\begin{align}\label{exercise}
 \cZ (\xi,\zeta)
 =& 1 + \xi\zeta \langle W| D ( 1 - \xi(\zeta D+E) )^{-1}| V \rangle
 = \frac{1}{ 1 - \frac{\xi\zeta(1+\xi)}{\beta(1+\xi\zeta)} },\\
 Z=&\frac{\beta(1-\alpha)}{\beta-\alpha-\alpha\beta}.
\end{align}
The square root vanishes from $Z$,
 and thus the critical line loses the straight line part,
 see Fig. \ref{fig:pd184}:
\begin{align}
 \alpha_c=\frac{\beta}{1+\beta}.
\end{align}
The average length and the average numbers of particles in the system and the queue are simplified as
\begin{align}
 \langle L \rangle
  = \frac{\alpha}{\beta-\alpha-\alpha\beta},\quad
 \langle N \rangle
  = \frac{\alpha(1-\alpha)}{\beta-\alpha-\alpha\beta},\quad
 \langle N_q \rangle
  = \frac{\alpha^2}{\beta(\beta-\alpha-\alpha\beta)}.
\end{align}
It is also easy to calculate
 the probability distributions
 of $L$ and $N$
 by expanding  $\cZ(\xi,1)$ and  $\cZ(\frac{\alpha}{1-\alpha},\zeta)$ 
 around $\xi=0$ and $\zeta=0$, respectively:
\begin{align}
\begin{split}
 & {\rm Prob} [\text{the length of the system}=L] \\
 &= \frac{1}{Z} \left(\frac{\alpha}{1-\alpha}\right)^L
\frac{1}{L!}\left.\frac{ \partial^L \cZ(\xi,1)}{ \partial \xi^L }
\right|_{\xi=0}
= \frac{\beta-\alpha\beta-\alpha}{\beta(1-\alpha)}
     \left(\frac{\alpha}{\beta(1-\alpha)}\right)^L,
\end{split}
\\
\begin{split}
 & {\rm Prob}[\text{the number of particles in the system}=N] \\
 & = \frac{1}{Z}  \frac{1}{N!} \left.
  \frac{\partial^N \cZ(\frac{\alpha}{1-\alpha},\zeta)}{\partial \zeta^N }
  \right|_{\zeta=0}
 \\
 &=
  \begin{cases}
 \displaystyle
 \frac{\beta-\alpha\beta-\alpha}{\beta(1-\alpha)} & N=0 , \\
 \displaystyle
 \frac{\beta-\alpha\beta-\alpha}
      {\beta(1-\beta+\alpha\beta)(1-\alpha)}
    \left(\frac{\alpha(1-\beta+\alpha\beta)}
    {\beta(1-\alpha)^2}\right)^N  & N\ge 1 .
  \end{cases}
\end{split}
\end{align}

\section{Conclusion} \label{conc}

In this paper, we introduced 
 an extension of the discrete-time M/M/1 queueing process
 with excluded-volume effect (exclusive queueing process).
We took the parallel-update scheme.
A stationary-state solution was obtained
 in a slightly arranged matrix product form
 for the parallel-update open TASEP.
We found that 
 the critical line which separates the parameter space
 into regions with and without the stationary state
 is given by the stationary current
 of the open TASEP in the maximal-current
 and high-density phases.
Particularly, we should note that 
  the entry rate cannot be greater than
  the maximal current of the TASEP
  if the model converges.
We calculated the average length of the system
 (the position of the leftmost particle)
 and the average number of particles in the system.
These results recover
 the stationary state, the critical line
 and the average values of
 the continuous-time version
 of the exclusive queueing process \cite{RefA},
 in the limit where the time interval $\Delta t\to0$.
In this sense,
 our discrete-time model is a generalization
 of the continuous-time model.
We calculated the number of particles in the queue as well.

When $p=1$,
 i.e. the bulk hopping is deterministic,
 the matrices and the vectors
 constructing the stationary-state solution
 have a two-dimensional representation.
The probability distributions of
 the length of the system and the average number of particles
 are both geometric in this case.

We leave calculations of the density profile
  and correlation functions in the stationary state
  as future works.
In the TASEP with the ordinary open boundary condition,
   a domain wall theory explains its phase diagram successfully
    \cite{RefKSKS}.
Investigating how our model converges to the stationary state
   or diverges
   with a similar argument will be also an interesting study.

\section*{Acknowledgements}
This work is supported
 by Global COE Program
  ``Education and Research Hub for Mathematics-for-Industry''
 and Grants-in-Aid
  (for Young Scientists (B) 22740106 and
   for JSPS Fellows 20$\cdot$10918) from JSPS
   and Ministry of Education, Culture, Sports, Science and Technology.

\appendix

\section{Reduction of the balance equation} \label{appa}

In this section, we derive Eqs. \eqref{11}-\eqref{100}. 
We calculate the action of $\cT$ on $P$
 under the assumption \eqref{ass1} and \eqref{ass2}.
Let us use a short hand notation $a=\frac{\alpha}{p(1-\alpha)}$.
For a configuration $11\tau$ with $|11\tau|=L$,
 the action of $\cT$ is calculated as

\begin{align}
\begin{split}
& Z\cT P(11\tau)  \\
=&Z\sum_{\tau'\in\cA_{11}(\tau)} W(11\tau'\to11\tau)P(11\tau') 
  +Z\sum_{\tau'\in\cA_{11}(\tau)} W(101\tau'\to11\tau) P(101\tau') \\
 &+Z\sum_{\tau'\in\cA_{11}(\tau)} W(1\tau'\to11\tau)P(1\tau') \\
=&\sum_{\tau'\in\cA_{11}(\tau)} W(11\tau'\to11\tau)a^Lf_L(11\tau') 
  +\sum_{\tau'\in\cA_{11}(\tau)}
     W(101\tau'\to11\tau)a^{L+1}f_{L+1}(101\tau') \\
 &+\sum_{\tau'\in\cA_{11}(\tau)}
     W(1\tau'\to11\tau)a^{L-1}f_{L-1}(1\tau') \\
=&
	a^L\Bigg(
 \sum_{\tau'\in\cA_{11}(\tau)}
    \frac{1}{1-\alpha}W(11\tau'\to11\tau) f_L(11\tau') \\
 &\quad\quad\quad\quad\quad\quad\quad\quad\quad
 +\sum_{\tau'\in\cA_{11}(\tau)}
    \frac{1}{\alpha}W(1\tau'\to11\tau) f_L(01\tau')
  \Bigg) ,
\end{split}
\end{align}
where we used
\begin{align}
 W(101\tau'\to11\tau)=pW(11\tau'\to11\tau)
                     =\frac{p(1-\alpha)}{\alpha}W(1\tau'\to11\tau)
\end{align}
and
\begin{align}
 f_{L+1}(101\tau')=f_L(11\tau')+pf_{L-1}(1\tau'),\quad
 pf_{L-1}(1\tau')=f_L(01\tau').
\end{align}
From $P(11\tau)=\cT P(11\tau)$, we get \eqref{11}.

For a configuration $101\tau$ with $|101\tau|=L$,
 the action of $\cT$ is calculated as
\begin{align}
\begin{split}
& Z \cT P(101\tau) \\
= 
 &Z\sum_{\tau'\in\cA_{101}(\tau)} W(101\tau'\to101\tau)P(101\tau')
  +Z\sum_{\tau'\in\cA_{101}(\tau)} W(110\tau'\to101\tau)P(110\tau') \\
 &+Z\sum_{\tau'\in\cA_{101}(\tau)} W(10\tau'\to101\tau)P(10\tau')
  +Z\sum_{\tau'\in\cA_{101}(\tau)} W(1010\tau'\to101\tau)P(1010\tau') \\
 &+Z\sum_{\tau'\in\cA_{101}(\tau)} W(1001\tau'\to101\tau)P(1001\tau') \\
=& \sum_{\tau'\in\cA_{101}(\tau)}
  W(101\tau'\to101\tau)a^Lf_{L}(101\tau')
 +\sum_{\tau'\in\cA_{101}(\tau)}
  W(110\tau'\to101\tau)a^Lf_{L}(110\tau') \\
 &+\sum_{\tau'\in\cA_{101}(\tau)}
  W(10\tau'\to101\tau)a^{L-1}f_{L-1}(10\tau')
  + \sum_{\tau'\in\cA_{101}(\tau)}
  W(1010\tau'\to101\tau)a^{L+1}f_{L+1}(1010\tau') \\
 &+ \sum_{\tau'\in\cA_{101}(\tau)}
  W(1001\tau'\to101\tau)a^{L+1}f_{L+1}(1001\tau') \\
=& a^L\Bigg(\sum_{\tau'\in\cA_{101}(\tau)}
  \frac{1}{1-\alpha}W(101\tau'\to101\tau)f_{L}(101\tau')
  +\sum_{\tau'\in\cA_{101}(\tau)}
  \frac{1}{1-\alpha}W(110\tau'\to101\tau)f_{L}(110\tau') \\
 &+\sum_{\tau'\in\cA_{101}(\tau)}
  \frac{1}{\alpha}W(10\tau'\to101\tau)f_{L}(010\tau')\Bigg),
\end{split}
\end{align}
where we used
\begin{align}
 W(1001\tau'\to101\tau)=&\frac{p}{1-p}W(101\tau'\to101\tau), \\
 W(1010\tau'\to101\tau)
  =& p W(110\tau'\to101\tau)=\frac{p(1-\alpha)}{\alpha} W(10\tau'\to101\tau),
\end{align}
and 
\begin{align}
 f_{L+1}(1010\tau')=f_L(110\tau')+pf_{L-1}(10\tau'),\\
 pf_{L-1}(10\tau')=f_L(010\tau'),\quad
 f_{L+1}(1001\tau')=(1-p)f_L(101\tau').
\end{align}
From $P(101\tau)=\cT P(101\tau)$, we get \eqref{101}.

For a configuration $100\tau$ with $|100\tau|=L$,
 the action of $\cT$ is calculated as
\begin{align}
\begin{split}
 & Z\cT P(100\tau) \\
=&Z\sum_{\tau'\in\cA_{100}(\tau)}
 W(10\tau'\to100\tau)P(10\tau')
 +Z\sum_{\tau'\in\cA_{100}(\tau)}
 W(100\tau'\to100\tau)P(100\tau') \\
=&\sum_{\tau'\in\cA_{100}(\tau)}
 W(10\tau'\to100\tau)a^Lf_L(10\tau')
 +\sum_{\tau'\in\cA_{100}(\tau)}
 W(100\tau'\to100\tau)a^{L+1}f_{L+1}(100\tau') \\
=& a^L\Bigg(
 \sum_{\tau'\in\cA_{100}(\tau)}
 \frac{1}{1-\alpha}W(10\tau'\to100\tau)f_L(10\tau') \\
 & +\sum_{\tau'\in\cA_{100}(\tau)}
   \frac{1}{(1-\alpha)(1-p)}W(10\tau'\to100\tau)f_L(00\tau')
 \Bigg),
\end{split}
\end{align}
where we used
\begin{align}
 W(100\tau'\to100\tau)=\frac{p}{1-p}W(10\tau'\to100\tau)
\end{align}
and 
\begin{align}
 f_{L+1}(100\tau')=(1-p)f_L(10\tau'),\quad
 f_L(00\tau')=0.
\end{align}
From $P(100\tau)=\cT P(100\tau)$, we get \eqref{100}.

\section{Derivation of $\cZ(\xi,\zeta)$} \label{appb}

According to \cite{RefERS},
 $D,E,\W{}$ and $\V{}$ can be reduced as
\begin{align}\label{reduceDE}
 D=\twotwo{D_1}{0}{D_2}{0},\quad
 E=\twotwo{E_1}{E_2}{0}{0},\quad
 \W{}=(\ \W{1}\quad \W{2}\ ),\quad
 \V{}=\left( \begin{array}{c} \V{1} \\ \V{2} \end{array}\right),
\end{align}
where matrices $D_1,D_2,E_1$ and $E_2$,
 and vectors $\W{1},\W{2},\V{1}$ and $\V{2}$ satisfy
\begin{align}
 D_1E_1=(1-p)(D_1+E_1+p), \label{D1E1}\\
 D_1\V{1}=\frac{p(1-\beta)}{\beta}\V{1},\quad
 \W{1}E_1=0,\label{D1V1W1E1}\\
 E_2D_2=p(D_1+E_1+p),
 \label{E2D2} \\
 E_2\V{2}=p\V{1},\quad
 \W{2}D_2=\W{1}p \label{E2V2}.
\end{align}
We can easily show that $D,E,\W{}$ and $\V{}$ satisfy
 the relations \eqref{algrel} from the relation
 \eqref{D1E1}-\eqref{E2V2}.
In view of the normalization \eqref{WDV},
 we impose $\W{1}V_1\rangle=1$.
The following gives a representation
 for these matrices and vectors,
 which will not be used in calculating $\cZ(\xi,\zeta)$:
\begin{align}
\begin{split}
\W{1}=(1,0,0,\dots),\quad
\W{2}=\sqrt{\beta}(0,1,-\sqrt{1-p},(-\sqrt{1-p})^2,\dots),\\
 D_1=
 \left(\begin{array}{ccccc}
  \frac{p(1-\beta)}{\beta} & \sqrt{\frac{p(1-p)}{\beta}} & \  &  \ &  \ \\
  \ & 1-p & \sqrt{1-p}  &  \ &  \  \\
  \  & \ & 1-p &  \sqrt{1-p} &  \  \\
  \  & \  & \ & 1-p &  \ddots  \\
  \  & \  & \  & \ & \ddots \ 
 \end{array}\right),
 \quad D_2=\sqrt{\frac{p}{1-p}}E_1,\\
 E_1=
 \left(\begin{array}{ccccc}
  0 & \ & \  &  \ &  \ \\
  \sqrt{\frac{p(1-p)}{\beta}} & 1-p & \  &  \ &  \  \\
  \  & \sqrt{1-p} & 1-p &  \ &  \  \\
  \  & \  & \sqrt{1-p} & 1-p &  \  \\
  \  & \  & \  & \ddots & \ddots \ 
 \end{array}\right),
 \quad  E_2=\sqrt{\frac{p}{1-p}}D_1,\\
\V{1}=\left(\begin{array}{c}1\\ 0\\ 0\\ \vdots\end{array}\right),\quad
\V{2}=\frac{\beta}{1-\beta}\sqrt{\frac{1-p}{p}}\V{1},
\end{split}
\end{align}
where we arranged the representation (6.6)-(6.11) in \cite{RefERS}.

To calculate $\cZ (\xi,\zeta)$,
we imitate the calculation \eqref{exercise} for $p=1$ as
\begin{align}
\begin{split}
 \cZ (\xi,\zeta)
 =& 1 + \sum_{L\ge 1} \xi^L\langle W| \zeta D (\zeta D+E)^{L-1} \V{}  \\
 =& 1 + \xi\zeta \W{} D ( 1 - \xi(\zeta D+E) )^{-1} \V{} \\
 =& 1 + \xi\zeta \left(\ \W{1}\quad \W{2}\ \right)
     \twotwo{D_1}{0}{D_2}{0}
      {\twotwo{1-\xi\zeta D_1-\xi E_1\ }{\ -\xi E_2}
      {-\xi\zeta D_2}{1}}^{-1}
      \left( \begin{array}{c} \V{1} \\ \V{2} \end{array}\right)\\
 =& 1 + \xi\zeta \left(\ \W{1}\quad \W{2}\ \right)
     \twotwo{D_1}{0}{D_2}{0}{\twotwo{X}{\ \xi X E_2}
      {\xi\zeta D_2 X\ }{\ 1+\xi^2\zeta D_2XE_2}}
      \left( \begin{array}{c} \V{1} \\ \V{2} \end{array}\right) \\
 =& 1 + \xi\zeta \left(\ \W{1}D_1+\W{2}D_2 \ \right)
      \left( X\V{1} + \xi X E_2\V{2} \right),
\end{split}
\end{align}
where $X^{-1}=1-\xi\zeta D_1-\xi E_1-\xi^2\zeta E_2D_2$.
Using the relation \eqref{E2D2},
 we can eliminate $D_2$ and $E_2$ in $X$ as
\begin{align}
 X^{-1}=(1-p^2\xi^2\zeta) - \xi\zeta(1+p\xi)D_1 - \xi(1+p\xi\zeta)E_1.
\end{align}
Furthermore, using the relation \eqref{E2V2},
 we can eliminate $D_2,E_2,\W{2}$ and $\V{2}$ in $\cZ(\xi,\zeta)$ as
\begin{align}
\begin{split}
 \cZ(\xi,\zeta) =& 1 + \xi\zeta \W{1}(D_1+p)X(1+p\xi)\V{1}\\
  =& 1 + \W{1}\left\{
      \xi\zeta(1+p\xi)D_1+p\xi\zeta(1+p\xi)\right\}X\V{1} \\
  =& 1 + \W{1}\left\{ -X^{-1}+(1+p\xi\zeta)-\xi(1+p\xi\zeta)E_1
      \right\}X\V{1} \\
  =& (1+p\xi\zeta)\W{1}X\V{1}.
\end{split}
\end{align}
In the last equality,
 we used the second relation of \eqref{D1V1W1E1}
 and $\langle W_{1}|V_{1}\rangle=1$.
Now we borrow an idea from Section 4.3 of \cite{RefBE},
 where the generating function of the TASEP with a single defect particle
 on a ring was obtained.
Set 
\begin{align}
 D'=aD_1+b,\quad E'=cE_1+d
\end{align}
with $a=\sqrt{\frac{u}{1-p}},b=1-\sqrt{u(1-p)},
c=\frac{1}{\sqrt{u(1-p)}}$ and $d=1-\sqrt{\frac{1-p}{u}}$,
and we get
\begin{align}
 D'E'&=D'+E' \\
 \W{1}E'&=d\W{1},\quad
 D'\V{1}=\left(\frac{p(1-\beta)}{\beta}a+b\right)\V{1}
 =:f\V{1},
\end{align}
noting the relations \eqref{D1E1} and \eqref{D1V1W1E1}.
Set $u=\frac{\zeta(1+p\xi)}{1+p\xi\zeta}$,
and we get
\begin{align}
\begin{split}
 X^{-1}
 =& (1-p^2\xi^2\zeta)-\xi(1+p\xi\zeta)(E_1+uD_1) \\
 =& g-h(D'+E'),
\end{split}
\end{align}
where
\begin{align}
g=(1-p^2\xi^2\zeta)+\xi(1+p\xi\zeta)\frac{b+d}{c}, \quad
h=\frac{\xi(1+p\xi\zeta)}{c}.
\end{align}
Set $\omega(1-\omega)=\frac{h}{g}$, and we get
\begin{align}
\begin{split}
 X=\frac{1}{g}\left(1-\frac{h}{g}(D'+E')\right)^{-1}
  =\frac{1}{g}\left(1-\omega E'\right)^{-1}
     \left(1-\omega D'\right)^{-1}.
\end{split}
\end{align}
Finally, we achieve 
\begin{align}
 \cZ(\xi,\zeta)
 =\frac{1+p\xi\zeta}{g}
   \W{1}\left(1-\omega E'\right)^{-1}
     \left(1-\omega D'\right)^{-1}\V{1}
 =\frac{1+p\xi\zeta}{g(1-\omega d)(1-\omega f)},
\end{align}
which can be simplified as \eqref{closedform}
and expanded as
\begin{align}\label{sum}
\cZ(\xi,\zeta)=&1+\sum_{L\ge N\ge 1}a_{LN}\xi^L\zeta^N , \\
\begin{split}
 a_{LN} =&
 (1-p)^{L-2N} \sum_{k=0}^{N}
 \frac{(L-k-1)!(N-1)!(-p)^k}{(L-N)!(N-k-1)!k!(N-k)!} \\
           &\!\!\!\!\!\!\!\!\!\!\!\!\!\!\!\!
 \times  \sum_{k=0}^{N-1} \left[
           \left(\frac{p}{\beta}\right)^{k+1} \frac{(L-k-2)!}{(N-k-1)!(L-N-1)!}
    \sum_{\ell=0}^{k}\frac{(k-\ell+1)\cdot N!}{\ell!(N-\ell)!}(-\beta)^\ell
              \right] .
\end{split}
\end{align}

\section{Derivation of $\langle N_q\rangle$}\label{appc}

In this section, we show
\begin{align}\label{forNq}
 \sum_{L\ge 2} \xi^L
  \W{} \zeta D (\zeta D+E)^{L-2} (D+E) \V{}
 = \left(1+\frac{p\xi(1-\zeta)}{\beta(1+p\xi\zeta)}\right)
    \cZ(\xi,\zeta)-\frac{\beta+p\xi}{\beta}.
\end{align}
The left-hand side is calculated as
\begin{align}
\begin{split}
  &\sum_{L\ge 2} \xi^L
  \W{} \zeta D (\zeta D+E)^{L-2} (D+E) \V{} \\
 =&\sum_{L\ge 1} \xi^L\W{} D (\zeta D+E)^{L-1}\V{} \\
  &-(1-\zeta)\sum_{L\ge 2} \xi^L\W{} D (\zeta D+E)^{L-2}E\V{}
   -\xi\W{} D\V{} \\
 =&\frac{\cZ(\xi,\zeta)-1}{\zeta}
    -\xi^2(1-\zeta)\W{} D (1-\xi(\zeta D+E))^{-1}E\V{}
   -\frac{p\xi}{\beta}
\end{split}
\end{align}
Recall the formula 
\begin{align}
(1-\xi(\zeta D+E))^{-1}=\twotwo{X}{\ \xi X E_2}
      {\xi\zeta D_2 X\ }{\ 1+\xi^2\zeta D_2XE_2}.
\end{align}
Using this and the relations
 \eqref{D1V1W1E1} and \eqref{E2V2}, we obtain
\begin{align}
\begin{split}
 &\W{} D (1-\xi(\zeta D+E))^{-1}E\V{} \\
 &=\W{1}(D_1+p)X(E_1+p)\V{1} \\
 &=\W{1}
   \frac{(1+p\xi\zeta)-\xi(1+p\xi\zeta)E_1-X^{-1}}{\xi\zeta(1+p\xi)}X
   \frac{(1+p\xi)-\xi\zeta(1+p\xi)D_1-X^{-1}}{\xi(1+p\xi\zeta)}
   \V{1} \\
 &=\frac{\W{1}\left\{(1+p\xi\zeta)-X^{-1}\right\}X
   \left\{(1+p\xi)\left(
    1-\frac{p(1-\beta)\xi\zeta}{\beta}\right)-X^{-1}\right\}
   \V{1}}{\xi^2\zeta(1+p\xi)(1+p\xi\zeta)}
\end{split}
\end{align}
Finally, noting
 $\W{1}X\V{1}=\frac{\cZ(\xi,\zeta)}{1+p\xi\zeta}$, $\W{1}V_1\rangle=1$
 and the relation \eqref{D1V1W1E1},
 we achieve \eqref{forNq}.

\bibliographystyle{spmpsci}      


\end{document}